\documentstyle[12pt]{article}
\newcommand{\SC}{\langle{\cal O}_8^{\psi '}(^1S_0)\rangle}
\newcommand{\SD}{\langle{\cal O}_8^{\psi '}(^3S_1)\rangle}
\newcommand{\PS}{\langle{\cal O}_8^{\psi '}(^3P_0)\rangle}

\begin{document}
\renewcommand{\thefootnote}{\fnsymbol{footnote}}
\begin{flushright}
KOBE--FHD--97--03\\
FUT--TH--97--02~~\\
HUE--26
June~~~~~~~~1997
\end{flushright}
\begin{center}
{\LARGE\bf {\boldmath $\psi '$} productions as a test of color--octet
mechanism}\\

\vspace{3.5em}

T. Morii, D. Roy and K. Sudoh\\
\vspace{0.8em}
{\it Faculty of Human Development and}\\
{\it Graduate School of Science and Technology,}\\
{\it Kobe University, Nada, Kobe 657, Japan}\\
\vspace{1em}
Zuxun Sun\\
\vspace{0.8em}
{\it China Institute of Atomic Energy,}\\
{\it Beijing 102413, P.O.Box 275, P. R. China}\\
\vspace{1em}
S. Tanaka\\
\vspace{0.8em}
{\it Department of Physics,}\\
{\it Hyogo University of Education, Yashiro--cho, Hyogo 673--14, Japan}\\
\vspace{1em}
and\\
\vspace{1em}
T. Yamanishi\\
\vspace{0.8em}
{\it Department of Management Science,}\\
{\it Fukui University of Technology, Gakuen, Fukui 910, Japan}\\
\vspace{4.5em}
{\bf Abstract}
\end{center}

\baselineskip=24pt

To test the color-octet model of heavy quarkonium productions, we propose
$\psi '$--productions at small--$p_T$ regions in polarized $pp$ 
collisions for the forthcoming
RHIC polarized experiments, whose 
experimental test at $\sqrt s=$50 GeV might be very promising.

\vspace{1.5em}

\vfill\eject

Traditionally, heavy quarkonium productions have been calculated
so far in the color--singlet model\cite{Schuler}, which 
is essentially a nonrelativistic model. However, with this model one cannot 
quantitatively estimate various uncertainties originated from 
the higher order QCD corrections,
the quarkonium binding effects and the corrections due to relativistic
effects of the quarkonium.
Recently it has been reported that the cross 
sections of prompt $J/\psi$ and
$\psi '$ productions in unpolarized $p\bar p$ collisions measured by
the CDF collaboration are largely 
inconsistent with the calculation by the QCD lowest
order process with the color--singlet mechanism 
alone\cite{CDF}.
This suggests that we need some other mechanisms beyond the 
color-singlet model.

In these years, a new color--octet model has
been advocated by several people\cite{octets}
as one of the most promising candidates that could 
remove such a big discrepancy between the experimental data and the 
prediction of the color--singlet model.   
The model is quite successful in 
explaining the CDF data for large--$p_T$ heavy quarkonium productions.
About the same time, a rigorous formulation of such a new model
has been presented in terms of a beautiful effective theory called the
nonrelativistic QCD(NRQCD), in which the ${\cal O}(v)$ corrections of the 
relative velocity between the bound heavy quarks can be systematically
calculated\cite{Bodwin}.
Physics of the color--octet model 
is now one of the most interesting topics for 
heavy quarkonium productions at high energy.  
Several processes have been already suggested for testing
the color-octet model, such as 
transversely polarized prompt J/$\psi$ and $\psi '$ hadroproductions at
high energy collisions\cite{Cho},
polar angle distributions of the J/$\psi$ in $e^+ e^-$ annihilations into
J/$\psi + X$\cite{Braaten},
Z$^0$ decays at LEP\cite{Cheung} and so on.
However, the prediction of the color-octet model on 
$\gamma + p\to J/\psi +X$ is at variance with recent data
at HERA\cite{Cacciari} and thus 
the discussion seems still controversial.  To go beyond the present
theoretical understandings, it is necessary to study other processes. 

In this paper, as another test of the color--octet model, we propose
a different process, the $\psi '$--hadroproduction at small--$p_T$
regions in longitudinally polarized proton--longitudinally polarized proton
collisions which will be observed in the forthcoming RHIC experiment.
The process is of great advantage to clearly test
the color-octet model as described in the following.  Since the process 
is dominated by the s--channel gluon--gluon fusion, there is no 
direct productions of the color-singlet $\psi '$
because of charge conjugation.
For this process, only two 
states are expected to contribute to the $\psi '$--production in the final
state: (1)a color--octet state, where a $c\bar c$ pair 
is produced at short distances in a color--octet state and subsequently
evolves nonperturbatively into a physical quarkonium\cite{octets},
(2)a radially excited color--singlet 2$^3$P$_2$ state($\approx$ 3.9 -- 4.0 GeV)
decaying into $\psi '$ + $\gamma$, where the decay 
into $D\bar D$, $D\bar D^{\ast}$ is suppressed by $D$--wave phase space and
dynamical effects\cite{LeYaouane}.
Contribution of the 2$^3$P$_0$ state is considered to be small because
the branching ratio of the 2$^3$P$_0$ into $\psi '$ + $\gamma$ is expected
to be very small by analogy of the tiny branching ratio of 1$^3$P$_0$ into
$J/\psi$ + $\gamma$, Br(1$^3$P$_0$$\rightarrow $$J/\psi$ + $\gamma$)
=(6.6$\pm 1.8$)$\times 10^{-3}$\cite{PDG}, and thus can be safely
neglected here.
Other charmonium states, 2$^{-+}$, 
2$^{--}$($\approx$ 3.81 to 3.85 GeV) and the 2$^3$P$_1$ state 
($\approx$ 3.9GeV), do not contribute to this process because of 
charge conjugation and Yang's theorem, respectively, though
these states might contribute to large--$p_T$ 
$\psi '$--productions\cite{Close95}.
Note that in the case of $J/\psi$ productions, in addition 
to radiative decays of 
1$^3$P$_2$ and $2^3$P$_2$ states, $\psi '$$\rightarrow $$J/\psi$ + $X$
contributes to the $J/\psi$ production in the final states and
hence the analysis must be complicated.
Furthermore, since the $\psi '$ is dominantly produced in
gluon fusion,
the cross section is sensitive to the 
gluon density in the proton and thus one can get 
a good information on the spin--dependent gluon distribution 
in the proton by analyzing this polarized process, which is also
a hot current topic.  Related subject has been studied recently by 
Teryaev and Tkabladze\cite{teryaev97}: they have calculated 
the two--spin asymmetry of the $J/\psi$ production at large $p_T$($>$1.5GeV)
regions in polarized $pp$ collisions and insisted that the color--octet
mechansim dominantly contributes to the asymmetry.

Let us introduce a two--spin asymmetry $A_{LL}$ for this process,
\begin{equation}
A_{LL} = \frac{\left[d\sigma_{++}-d\sigma_{+-}+
d\sigma_{--}-d\sigma_{-+}\right]}
{\left[d\sigma_{++}+d\sigma_{+-}+
d\sigma_{--}+d\sigma_{-+}\right]} = \frac{d\Delta\sigma}{d\sigma}~,
\label{eqn:A_LL}
\end{equation}
where $d\sigma_{+-}$, for instance, denotes that the helicity
of one beam particle is positive and the other is negative.

The spin--dependent and spin--independent differential cross sections of 
small--$p_{_T}$ $\psi '$--productions via the color--octet
state are given by\cite{Gupta}
\begin{eqnarray}
\frac{d\Delta\sigma_{CO}}{dx_{_L}}&=&
\frac{d\sigma_{++}}{dx_{_L}} - \frac{d\sigma_{+-}}{dx_{_L}} +
\frac{d\sigma_{--}}{dx_{_L}} - \frac{d\sigma_{-+}}{dx_{_L}} \nonumber\\
&=&\frac{\tau_{c}}{\sqrt{x_{_L}^2+4\tau_{c}}}\left[\frac{\pi^3\alpha_s^2}
{144m_c^5}
\{\SC-\frac{1}{m_c^2}\PS\}
\Delta g(x_a, Q^2)\Delta g(x_b, Q^2)\right .\nonumber\\
&&\left .-\frac{\pi^3\alpha_s^2}{54m_c^5}\SD
\{\Delta q(x_a, Q^2)\Delta\bar q(x_b, Q^2)+
\Delta q\leftrightarrow\Delta\bar q\} \right]~,
\end{eqnarray}
\begin{eqnarray}
\frac{d\sigma_{CO}}{dx_L}&=&
\frac{\tau _c}{\sqrt{x_L^2+4\tau _c}}\left[\frac{\pi^3\alpha_s^2}{144m_c^5}
\{\SC+\frac{7}{m_c^2}\PS\} g(x_a, Q^2)~g(x_b, Q^2)\right .\nonumber\\
&&\left .+\frac{\pi^3\alpha_s^2}{54m_c^5}\SD
\{q(x_a, Q^2)~\bar q(x_b, Q^2)+q\leftrightarrow\bar q\}\right]~,
\label{eqn:dDs-co}
\end{eqnarray}
where $x_a$ and $x_b$ are the momentum fraction in a proton and given as
\begin{equation}
x_a=\frac{x_L+\sqrt{x_L^2+4\tau _c}}{2}~,~~
x_b=\frac{-x_L+\sqrt{x_L^2+4\tau _c}}{2}~,~~
x_L\equiv \frac{2p_L}{\sqrt s}~,~~
\tau _c\equiv \frac{4m_c^2}{s}~,
\label{eqn:def-x}
\end{equation}
with longitudinal momentum $p_L$ of the produced particle.
$\Delta g(x, Q^2)$ and $\Delta q(x, Q^2)$ are the 
spin--dependent gluon and quark density with the momentum
fraction $x$ at any $Q^2$, respectively.
$\SC$, $\SD$ and $\PS$ are nonperturbative long--distance factors
associated with the production of a $c\bar c$ pair in a color--octet
$^1S_0$, $^3S_1$ and $^3P_0$ states, respectively.
From recent analysis on charmonium hadroproductions, the value of $\SD$
and of the combination are given as
$\SD\approx 4.6\times 10^{-3}~{\rm [GeV^3]}$,
$\SC+\frac{7}{m_c^2}\PS\approx 5.2
\times 10^{-3}~{\rm [GeV^3]}$ \cite{Beneke96},
and another combination
$\frac{1}{3}\SC+\frac{1}{m_c^2}\PS\approx (5.9\pm 1.9)\times
10^{-3}~{\rm [GeV^3]}$ \cite{Leibovich96}.
Then we find the ratio as
$\frac{\tilde\Theta}{\Theta}\equiv
\frac{\SC-\frac{1}{m_c^2}\PS}{\SC+\frac{7}{m_c^2}\PS}
\approx 8.0 - 3.6$.  Using various parton distributions, we
have numerically calculated the $A_{LL}$ due to the color--octet state
alone and found that it becomes positive in all $x_L$ region if
$\Delta g(x, Q^2)$ does not change its sign in $0<x<1.0$, though
its value largely depends on $\frac{\tilde\Theta}{\Theta}$.

The spin--dependent differential cross section of 
small--$p_{_T}$ $\psi '$--productions via 
the 2$^3$P$_2$ state can be given by\cite{Morii96}  
\begin{eqnarray}
\frac{d\Delta\sigma_{2^3P_2}}{dx_{_L}}&=&
\frac{d\sigma_{++}}{dx_{_L}} - \frac{d\sigma_{+-}}{dx_{_L}} +
\frac{d\sigma_{--}}{dx_{_L}} - \frac{d\sigma_{-+}}{dx_{_L}} \nonumber\\
&=&Br(2^3P_2\to\psi '+\gamma) \nonumber\\
&&\times\frac{-16\pi^2\alpha_s^2|R_{2^3P_2}'(0)|^2}{M^7}
\frac{\tau}{\sqrt{x_{_L}^2+4\tau}}
\Delta g(x_a, Q^2)~\Delta g(x_b, Q^2)~,
\label{eqn:dDs-chi}
\end{eqnarray}
where $x_a$ and $x_b$ are given by replacing $\tau_{c}$ 
in eq.(\ref{eqn:def-x}) by $\tau\equiv \frac{M^2}{s}$. The spin--independent
cross section is given by replacing $\Delta g(x, Q^2)$ by $g(x, Q^2)$.
As shown from eq.(\ref{eqn:dDs-chi}), the $A_{LL}$ due to the 
radiative decay of the 2$^3$P$_2$ state
becomes negative, though its magnitude
depends on $|R_{2^3P_2}'(0)|$ whose value has been calculated 
using various potentials models as
$|R_{2^3P_2}'(0)|^2=$0.076, 0.102, 0.131, and 0.186GeV$^5$, for
the logarithmic, Buchm\"uller-Tye, power-low, and Cornell potentials,
respectively\cite{Eichten}.

Relativistic Heavy Ion Collider (RHIC), which is designed to
have a luminosity of 2$\times$10$^{32}$ cm$^{-2}$sec$^{-1}$, an energy of
$\sqrt s=50-500$ GeV and a beam polarization of about 70 \%,
is now under construction at Brookhaven National Laboratory (BNL)
and, hopefully in a few years, comes to produce 
fruitful data on high energy polarized $pp$ collisions.
Expecting the forthcoming RHIC experiments, we have calculated
the $A_{LL}$ for these energies.
In this calculation, we need an information of $\Delta g(x)$
and, in addition, $\Delta q(x)$ and $\Delta \bar q(x)$ 
for the case of the color--octet model.
So far, many people have suggested various kinds of different spin--dependent
gluon distributions from the analysis of data on nucleon spin structure
functions\cite{GS95,BBS95,GRV95}.
Among many models of $\Delta g(x, Q^2)$,
we take here typical three types of $\Delta g(x, Q^2)$, (a)set A of 
GS parametrization\cite{GS95}, (b)the BBS parametrization\cite{BBS95}, 
and (c) the GRV parametrization\cite{GRV95}, which are shown in fig. 1, and  
calculate the $A_{LL}$ for $\psi '$--productons 
as a function of $x_{_L}$ at RHIC energies.
In calculating for the color--octet model, we have taken
$\Delta q(x)$ and $\Delta \bar q(x)$ from respective models of
GS\cite{GS95}, BBS\cite{BBS95}, and GRV\cite{GRV95}, and found that 
the contribution of $q\bar q\rightarrow c\bar c$ 
is considerably smaller than the one of
$gg\rightarrow c\bar c$ for $x_L<0.5$. 
As for the spin--independent parton distributions, we have used
the Owens parametrization
\cite{Owens91} for (a), the BBS parametrizaton \cite{BBS95} for (b), and 
the GRV parametrization\cite{GRVogt95} for (c).

Fixing $Q^2$ as $4m_c^2$ with 
$m_c=1.5$GeV and taking the mass
M$_{2^3P_2}$=3.98GeV and Br(2$^3$P$_2\to\psi '+\gamma$)=0.08\cite{Roy}, 
we have studied the parameter dependence of 
$A_{LL}=\frac{d\Delta\sigma_{CO}/dx_{_L}+d\Delta\sigma_{2^3P_2}/dx_{_L}}
{d\sigma_{CO}/dx_{_L}+d\sigma_{2^3P_2}/dx_{_L}}$
on $\frac{\tilde\Theta}{\Theta}$ and $|R_{2^3P_2}'(0)|^2$ at relevant
RHIC energies 
and found that the $A_{LL}$ becomes larger for
the larger $\frac{\tilde\Theta}{\Theta}$ and 
smaller $|R_{2^3P_2}'(0)|^2$.  
We found that the $A_{LL}$ became positive in all regions of 
$\frac{\tilde\Theta}{\Theta}$ and $|R_{2^3P_2}'(0)|^2$ given above
for $\sqrt s=50 - 500$GeV.
The $A_{LL}$ is also largely dependent of
$\Delta g(x, Q^2)$.
The results calculated at $\sqrt s=50$GeV for two extreme cases, 
(A)$\frac{\tilde\Theta}{\Theta}=8.0$, $|R_{2^3P_2}'(0)|^2=0.076$GeV$^5$
and (B)$\frac{\tilde\Theta}{\Theta}=3.6$, $|R_{2^3P_2}'(0)|^2=0.186$GeV$^5$,
are presented in fig. 2.  Since the $A_{LL}$ is rather large, it 
must be easy to test the color-octet model in the future experiment.
To see the energy dependence, we have calculated the $A_{LL}$ at the
highest RHIC energy,
$\sqrt s=500$GeV, which becomes very small as shown in fig. 3.
This is due to the fact that at larger $\sqrt s$, $x_a$ and $x_b$ ($=x_a-x_L$)
defined by eq.(\ref{eqn:def-x}) take smaller value and hence 
$\Delta g(x_a)/g(x_a)\times\Delta g(x_b)/g(x_b)$ becomes smaller.
Now, it is important to note that without color--octet contributions, 
we can never expect a positive $A_{LL}$.
Therefore, if we observe a positive $A_{LL}$ in the forthcoming
RHIC experiment, we can definitely say that the color--octet 
mechanism really contributes to this process.
Our results suggest that the observation of the $A_{LL}$
is very effective for testing the color--octet model and, 
in practice, the experiment at $\sqrt s=50$GeV is very promising.
Since the results significantly depend on the value of 
$\frac{\tilde\Theta}{\Theta}$, it is very important to 
determine the value from other experiments to give
a better prediction.

In summary, we have proposed the small--$p_T$ $\psi '$--production 
in longitudinal proton--longitudinal proton collisions whose 
experimental test will be available in the forthcoming RHIC experiments.
The process can be dominated by two--gluon fusion in
the lowest order and thus we have only two mechanisms, i.e. 
color--octet and 2$^3$P$_2$ state
productions.  Since each of them
shows distinct behavior of the $A_{LL}$ with opposite
signs between the color--octet and 2$^3$P$_2$ state,
the process allows us to give a clean test of the color--octet
model.  Practically, the experimental test at $\sqrt s=50$GeV
might be most promising.  
Furthermore, the process is effective for testing
the spin--dependent gluon distribution in the proton 
because its cross section is directly proportional
to the product of $\Delta g(x)$ in both protons.

One of the authors(Z.S.) would like to thank the radiation group
of RIKEN for their kind hospitality.  T.Y. would like to thank the 
members of RCNP for allowing him to use the high performance computer
at RCNP.

\begin{center}
{\large \bf Figure captions}
\end{center}
\begin{description}

\item[Fig.\ 1:] The $x$ dependence of $x\Delta g(x, Q^2)$
at $Q^2=10$GeV$^2$ for various types of spin--dependent gluon distributions.
The solid, dashed and dotted lines indicate set A of
ref.\cite{GS95}, ref.\cite{BBS95} and ref.\cite{GRV95}, respectively.

\vspace{2em}

\item[Fig.\ 2:] The two--spin asymmetry $A_{LL}^{\psi '}(pp)$
for (A)$\frac{\tilde\Theta}{\Theta}=8.0$, $|R_{2^3P_2}'(0)|^2=0.076$GeV$^5$ 
and (B)$\frac{\tilde\Theta}{\Theta}=3.6$, $|R_{2^3P_2}'(0)|^2=0.186$GeV$^5$  
at $\sqrt s=50$GeV, calculated with various
types of $\Delta g(x)$, as a function of longitudinal momentum fraction $x_L$
of $\psi '$. Various lines represent the same as in fig.1.

\vspace{2em}

\item[Fig.\ 3:] The two--spin asymmetry $A_{LL}^{\psi '}(pp)$
for $\frac{\tilde\Theta}{\Theta}=8.0$ and $|R_{2^3P_2}'(0)|^2=0.076$GeV$^5$
at $\sqrt s=500$GeV. Various lines represent the same as in fig.1.
\end{description}
\end{document}